\input{aipcheck}

\documentclass[
    ,final            
  ]
  {aipproc}

\layoutstyle{6x9}

\begin{document}
\def\pim{$\pm$} 
\def\be{\begin{equation}}
\def\ee{\end{equation}}
\def\mdot{$\dot{m}$ }
\def\mpc{\,{\rm {Mpc}}}
\def\kpc{\,{\rm {kpc}}}
\def\kms{\,{\rm {km\, s^{-1}}}}
\def\msun{{$\rm M_{\odot}$}}
\def\Gyr{{\,\rm Gyr}}
\def\erg{{\rm erg}}
\def\sr{{\rm sr}}
\def\hz{{\rm Hz}}
\def\cm{{\rm cm}}
\def\sec{{\rm s}}
\def\eV{{\rm \ eV}}
\def\ledd{$L_{Edd}$~}
\def\mic{$\mu$ }
\def\ang{\AA }  
\def\cm2{cm$^2$ }
\def\se1{s$^{-1}$ }

\def\arcmin{\hbox{$^\prime$}}
\def\arcsec{\hbox{$^{\prime\prime}$}}
\def\degree{$^{\circ}$} 
\def\mic{$\mu$ }
\def\ang{\AA }  
\def\cm2{cm$^2$ }
\def\se{s$^{-1}$ }

\def\gtsima{$\; \buildrel > \over \sim \;$}
\def\ltsima{$\; \buildrel < \over \sim \;$}
\def\prosima{$\; \buildrel \propto \over \sim \;$}
\def\gsim{\lower.5ex\hbox{\gtsima}}
\def\lsim{\lower.5ex\hbox{\ltsima}}
\def\simgt{\lower.5ex\hbox{\gtsima}}
\def\simlt{\lower.5ex\hbox{\ltsima}}
\def\simpr{\lower.5ex\hbox{\prosima}}
\def\la{\lsim}
\def\ga{\gsim}
\def\Lsun{\rm L_{\odot}}
\def\Rsun{\rm R_{\odot}} 
\def\sr{A0620--00~}
\def\gx{GX~339$-$4~}

\def\ie{{\frenchspacing\it i.e. }}
\def\eg{{\frenchspacing\it e.g. }}
\def\etal{{~et al.~}}
\def\cxo{{Chandra~}}
\def\xmm{{XMM~}}
\title{Jets from the faintest black holes}

\author{Elena Gallo}{
  address={Physics Department, University of California, Santa Barbara, CA
  93106-9530, USA},altaddress={Chandra Fellow} }

\begin{abstract}
The question whether quiescent black hole X-ray binaries are capable of
powering relativistic outflows is addressed by means of simultaneous
radio/X-ray observations of a nearby system steadily emitting X-rays below
$10^{-8}$ times the Eddington luminosity. The robust detection of a radio
counterpart suggests that a synchrotron emitting outflow is being powered by
this system, even though its degree of collimation remains unknown, and hard
to investigate. With the inclusion of the A0620--00 data, the non linear
radio/X-ray correlation for hard state black hole X-ray binaries appears to
hold down to very low quiescent luminosities.  However, an increasing number
of outliers is being found at higher luminosities, questioning the
universality of such correlation, or at least its normalization.
\end{abstract}

\maketitle


\section{Quiescent accretion}
Among the various issues that remain open in the field of accretion onto black
hole binaries (BHBs) as well as super-massive black holes is the way the gas
accretes at very low Eddington ratios, in the so called {\it quiescent} regime
(where, for the binaries, the boundary between quiescence and a more active
regime can be set around $10^{33.5}$ erg \se, corresponding to a few
$10^{-6}$\ledd for a 10 \msun~BH; the reader is referred to McClintock \&
Remillard \cite{mccr06} for a review of X-ray states).  While there is little
doubt that the thin disc model~\cite{ss73} captures the basic physical
properties of BHBs in the thermal dominant state, the accretion mode
responsible for powering quiescent BHs is still a matter of debate.
Observations of highly sub-Eddington systems, most notably the Galactic Center
super-massive BH, paved the way to radiatively inefficient accretion flow
models (RIAFs).  By reviewing the vast literature on the subject, one
immediately comes to the conclusion that the most widely accepted/adopted
model for reproducing the spectral energy distribution of quiescent BHs is the
`advection-dominated accretion flow' solution (ADAF~\cite{ny94}\cite{ny95});
in this low-density, two temperature inflow, a significant fraction of the
viscously dissipated energy remains locked up in the ions as heat, and is
advected inward.  The ADAF model can successfully account for the overall
shape of the UV-optical-X-ray spectra of quiescent BHBs (see McClintock \etal
2003~\cite{mcc03} for an application to the high quality data of XTE
J1118+480).  Nevertheless, alternative suggestions are worth being
considered. For instance, convection, rather than advection, could be
responsible for keeping the plasma circulating in eddies along the flow by
transporting angular momentum inward, yielding zero net accretion rate
(CDAF;~\cite{quataert}).  A different solution is that elaborated by Blandford
\& Begelman \cite{bb99}, where the excess energy and angular momentum is lost
to an outflow at all radii; the final accretion rate into the hole may be only
a tiny fraction of the mass supply at large radii (ADIOS: `adiabatic
inflow-outflow solution'). Radiatively inefficient outflows may also be
responsible for dissipating the bulk of the liberated accretion power
(e.g.~\cite{mff},\cite{mnw05},\cite{yuan}).  Roughly speaking, the question
remains open whether quiescent BHBs are dim because of a highly reduced
radiative efficiency, or a highly reduced inner accretion rate.

Recent Chandra observations of nearby elliptical
galaxies~\cite{allen06} led to the discovery of a tight correlation
between the Bondi accretion rate, $\dot m_{Bondi}$ (as inferred from the
measured temperature, BH mass and density profile) and the power
emerging from these systems in the form of relativistic jets
(estimated by the work they exert on the observed X-ray cavities).  The
jets' power is found to be comparable to $0.1 \dot m_{Bondi}c^2$,
implying that a significant fraction of the matter entering the
accretion radius flows down to regions close to the black holes, where
the jets are thought to be formed. 

\section{quiescent jets?}
In the context of X-ray binaries, as well as super-massive black holes, the
term `jet' is typically used as a synonymous for a relativistic outflow of
plasma and implies a high degree of collimation (see Fender
2006~\cite{fender06} for a comprehensive review on X-ray binary jets). As a
matter of fact, high spatial resolution radio observations of BHBs in the hard
state \cite{mccr06} have imaged highly collimated structures in two systems
only: Cyg X-1~\cite{stirling} and GRS~1915+105~\cite{dhawan} are both resolved
into elongated radio sources on milliarcsec scales -- that is tens of A.U. --
implying collimation angles smaller than a few degrees on much larger scales
than the orbital separation.  Both systems display relatively high X-ray (and
radio) luminosities, with GRS 1915+105 being persistently close to the
Eddington limit (e.g.~\cite{fb04}), and Cyg X-1 having a bolometric X-ray
luminosity around 2 per cent of $ L_{Edd}$~\cite{disalvo01}.  This, however,
should not be taken as evidence against collimated jets at lower luminosities:
because of sensitivity limitations on current high resolution radio arrays,
resolving a radio jet at microJy level (if any) simply constitutes an
observational challenge. In addition, at such low levels, the radio flux could
be easily contaminated by synchrotron emission from the donor star.

The presence of a collimated outflow can
also be inferred by its long-term action on the local interstellar medium, as
in the case of the hard state BHBs 1E1740.7$-$2942 and GRS~1758$-$258, both
associated with arcmin-scale radio lobes~\cite{mirabel92},~\cite{marti02}.
Further indications come from the stability in the
orientation of the electric vector in the radio polarization maps of
GX~339$-$4 over a 
two year period~\cite{corbel00}. This constant position angle, being the
same as the sky position angle of the large-scale, optically thin radio jet
powered by GX 339$-$4 after its 2002 outburst~\cite{gallo04}, clearly
indicates a favoured ejection axis in the system.

On the other hand, failure to image a collimated structure in the hard
state of XTE J1118+480 down to a synthesized beam of 0.6$\times$1.0
mas$^{2}$ at 8.4 GHz~\cite{mirabel01} can challenge the {collimated}
jet interpretation, even though XTE J1118+408 was observed at roughly
one order of magnitude lower luminosity with respect to e.g. Cyg
X-1. If the jet size scaled as the radiated power, one could expect
the jet of XTE J1118+408 to be roughly ten times smaller than that of
Cyg X-1 (which is 2$\times$6 mas$^{2}$ at 9 GHz, at about the same
distance), \ie still point-like in the VLBA maps presented by
Mirabel \etal 2001~\cite{mirabel01}.

Garcia \etal 2003~\cite{garcia03} have pointed out that long period
($\simgt 1$ day) BHBs undergoing outbursts tend to be associated with
spatially resolved optically thin radio ejections, while short period
systems would be associated with unresolved, and hence physically
smaller, radio ejections. If a common production mechanism is at work
in optically thick and optically thin BHB jets~\cite{fbg}, then the
above arguments should apply to steady optically thick jets as well,
providing an alternative explanation to the unresolved radio emission
of XTE J1118+480, with its 4 hour orbital period, the shortest known
for a BHB. It is worth mentioning that, by analogy, a long period
system, like for instance V404 Cyg, might be expected to have a more
extended optically thick jet (approved high spatial resolution
observations of this system, with the High Sensitivity Array, will
hopefully answer this question).

\section{Radio/X-ray correlation in hard state (2003)}

In an attempt to assess the relation between accretion and jet
production in hard state BHBs, Corbel \etal 2003, and Gallo, Fender \&
Pooley 2003 (\cite{corbel03},\cite{gfp}) have established the
existence of a tight correlation between the X-ray and the radio
luminosity ($L_X$ and $L_{R}$), of the form $L_{ R}\propto
L_{X}^{0.7\pm 0.1}$. In those works, 10 hard state systems with nearly
simultaneous radio/X-ray observations were considered; the correlation
was found to hold over more than 3 orders of magnitude in $L_{X}$, up
to a few per cent of $L_{Edd}$, above which the sources enter the
thermal dominant state, and the core radio emission drops below
detectable levels.  Probably the most notable implication of the
non-linear scaling was the predicted existence of a critical X-ray
luminosity below which a significant fraction of the liberated
accretion power is channelled into a radiatively inefficient outflow,
rather than being dissipated locally by the inflow of gas and emitted
in the form of X-rays (this does not necessarily imply the X-ray
spectrum is dominated by non-thermal emission from the jet, as most of
the jet power may be stored as kinetic energy~\cite{fgj}).

\section{Radio and X-ray emission in quiescence: the case of A0620--00}

Due to its low X-ray luminosity ($L_X/L_{Edd}\sim10^{-8.5}$~\cite{kong})
and relative proximity, the BHB in 
\sr represented the most suitable known system to probe 
the radio/X-ray correlation beyond the hard state.  Indeed, deep VLA
observations, performed in 2005 August, resulted in the first radio
detection of a quiescent BHB emitting at such low X-ray luminosities;
the level of radio emission -- 51 $\mu$Jy at 8.5 GHz -- is the lowest
ever measured in an X-ray binary. At a distance of 1.2 kpc, this
corresponds to a radio luminosity $L_{\rm R}=7.5\times 10^{26}$ erg
sec$^{-1}$. By analogy with higher luminosity systems, partially
self-absorbed synchrotron emission from a relativistic outflow appears
as the most likely interpretation. Free-free wind emission is ruled
out on the basis that far too high mass loss rates would be required,
either from the companion star or the accretion disc, to produce
observable emission at radio wavelengths, while gyrosynchrotron
radiation from the corona of the companion star is likely to
contribute to less than 5 per cent to the measured flux density.

A0620--00 was observed simultaneously in the X-ray band with \cxo, its 
0.3-8 keV spectrum well fitted by an absorbed power law with photon
index $\Gamma=2.08^{+0.49}_{-0.35}$ and hydrogen equivalent column
density consistent with the optical value. The corresponding 2--10 keV
luminosity was about $7 \times 10^{30}$ erg sec$^{-1}$, a factor
of two higher than in a previous \cxo observation, in February
2000. 

\begin{figure}
\label{fig:1}
  \includegraphics[height=.6\textheight]{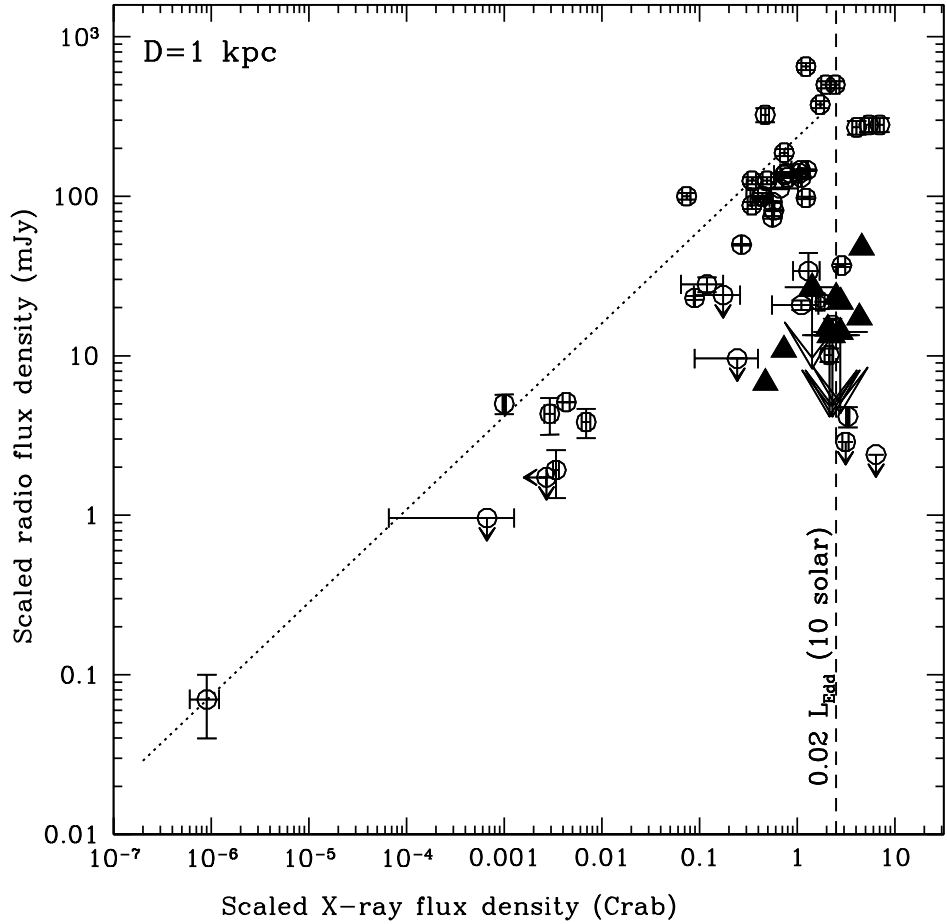} \caption{The
  radio/X-ray correlation in hard state BHBs~\cite{gfp}, with the
  addition of the $10^{-8.5}L_{Edd}$ A0620--00~\cite{gallo06} (open 
  circles). While this low Eddington ratio system appears to confirm
  the validity of a non-linear radio/X-ray correlation over more than
  6 orders of magnitude in $L_X$, an increasing numbers of hard state
  outliers is being found at higher luminosities (filled triangles),
  whose radio flux is typically lower with respect to the best-fitting
  power-law (e.g. XTE J1720--318~\cite{chaty},\cite{brock} and SWIFT
  J1753.5--0127~\cite{cadolle}, IGR J17497--2821~\cite{rodriguezj},
  XTE J1650-500~\cite{corbel04}). This challenges the validity of such
  relation, or at least the universality of its normalization.}
\end{figure}

The
simultaneous \cxo observations of \sr allowed us to test and extend
the radio/X-ray correlation for BHBs by 3 orders of magnitude in
$L_{\rm X}$.  The measured radio/X-ray fluxes seem to confirm the
existence of a non-linear scaling between the radio and X-ray
luminosity in this systems; with the addition of the \sr point
{$L_{\rm R}\propto L_{\rm X}^{0.58\pm0.16}$ provides a good fit to the
data for $L_{\rm X}$ spanning between $10^{-8.5}$ and $10^{-2} L_{\rm
Edd}$ (see Figure~\ref{fig:1}). The fitted slope, albeit consistent with the previously
reported value of $0.7\pm0.1$, is admittedly affected by the
uncertainties in the distance to GX339--4, for which the correlation
extends over 3 orders of magnitude in $L_{\rm X}$ and holds over
different epochs.

\section{Is the correlation truly Universal? }  

Even though the A0620--00 data appear to follow the hard state
radio/X-ray correlation, since 2003, when the compilation of
quasi-simultaneous radio/X-ray observations of hard state BHBs was
presented~\cite{gfp}, many outliers have been found. To mention a few
(those that came or were brought to the attention of this author at
least): XTE J1720--318~\cite{chaty},\cite{brock}, SWIFT
J1753.5--0127~\cite{cadolle}, IGR J17497--2821~\cite{rodriguezj} and
XTE J1650--500~\cite{corbel04}, while in the hard state, all appear to
lie significantly below the best-fitting correlation
(Figure~\ref{fig:1}).

{While a number of plausible ad-hoc `reasons' can be adduced on a
source-by-source basis, it simply seems more reasonable to conclude
that the radio/X-ray correlation may not be universal, or, at least,
that there is no universal normalization. This casts 
doubts on the possibility of relying on the best-fitting relation for
estimating other quantities, such as distance or BH mass.}  
A global re-assessment of the radio/X-ray correlation will be
presented in a forthcoming paper.
 
\section{Inflow vs outflow: energy budget}

While ADAF models predict the existence of bipolar outflows emanating
from the surface layers of the equatorial inflow~\cite{ny95},
generally they do not address the importance of such outflows with
respect to the overall accretion process in terms of energetics.
Ideally, one would like to be able to compare the total power carried
away in the form of outflows/jets, with the total accretion power
budget available to the black hole. As mentioned in the first Section,
this has been done for a handful of nearby elliptical
galaxies~\cite{allen06}; there, the mechanical jets power is found to
be comparable to $0.1 \dot m_{Bondi} c^2$. This has been interpreted
as evidence that a substantial fraction of the captured mass does
reach the innermost regions of the flow before being ejected in the
form of an outflow.\\

A similar conclusion has been reached in the case of the stellar mass
BH in A0620--00~\cite{gallo06}. 
Based on models for the optical/UV emission of the outer accretion
disc in dwarf novae~\cite{warner}, corrected downward to account for
the mass difference, McClintock \etal 1995~\cite{mcc95} estimate $\dot
M_{\rm out} = y 10^{-10} M_{\odot}$ yr$^{-1}$ for A0620--00, where $y$
is a factor of the order unity, that can be up to a few.  This value
is also comparable to the $3\times 10^{-11}$ \msun yr$^{-1}$ inferred
from the measure of the total energy released during the 1975 outburst
of \sr adopting 58 year recurrence time based on plate archives which
showed an outburst in 1917. This time-averaged value had been
calculated based on a distance of 1 kpc for \sr (vs. a refined value
of 1.2 kpc) and could still be underestimated by a factor 2 or so, to
allow for the possibility that an intermediate outburst was missed.
The putative luminosity associated with $\dot M_{\rm out}$, {if it was
to reach the black hole with standard radiative efficiency}, would be
$ L_{\rm tot} \equiv \eta \dot M_{\rm out}c^2 \simeq 6 \times 10^{35}
y~(\eta / 0.1)$ erg sec$^{-1}$, five orders of magnitude than the
observed X-ray (or bolometric) luminosity.  In the above expression
$\eta$ is the accretion efficiency, which depends only on the BH spin.
The various RIAFs provide different explanations for the much lower
luminosities that are observed in terms of different `sinks' for the
energy.

In the {ADAF} scenario, it is assumed that all the $\dot M_{\rm out}$ is
accreted onto the black hole ($\dot M_{\rm in}=\dot M_{\rm out}$)
while the radiated luminosity 
$L_{\rm bol}=\epsilon_{\rm rad} \dot M_{\rm in}c^2$ 
is much smaller
as a result of a reduced {radiative efficiency}
\begin{equation}
\label{eq:radeff}
\epsilon_{\rm rad} \equiv \eta f(\alpha) =\eta \times \left\{
        \begin{array}{ll}
        1,   &  \dot M_{\rm out} \ge \dot M_{\rm cr}   \\
        (\dot M_{\rm out}/\dot M_{\rm cr})^{\alpha}, & 
\dot M_{\rm out} < \dot M_{\rm cr}   \\
        \end{array}\right.\;
\end{equation}
where $\dot M_{\rm cr}$ is the critical rate above which the disc
becomes radiatively efficient. The index $\alpha$ is typically close
to unity, but its exact value may depend on the micro-physics of ADAF
and on how the bolometric luminosity is calculated.  Writing the
bolometric luminosity of \sr as $L_{\rm bol}= w~10^{32}$ erg
sec$^{-1}$, being $w$ a multiplicative factor, then
$f (\alpha)= \left(\frac{\dot M_{\rm out}}{\dot M_{\rm cr}}\right)^{\alpha} \simeq 1.7
\times 
10^{-4}~(w/y)~(0.1/\eta)$.  For $\alpha=1.3$,
$
\dot M_{\rm cr}\simeq 8\times 10^{-8}~y^{(1+1/1.3)}~w^{(-1/1.3)}~(\eta/0.1)^{(1/1.3)} ~M_{\odot}$ yr$^{-1}$,
which corresponds to an Eddington-scaled critical accretion rate $\dot
m_{\rm cr} \simeq 0.36~y^{(1+1/1.3)} w^{(-1/1.3)}
(\eta/0.1)^{(1+1/1.3)}$.  With $w\sim 5-10$, as implied by the ADAF
spectral modelling, a self-consistent ADAF solution is obtained for
$\dot m_{\rm cr}$ of a few times $10^{-2}$, as expected on theoretical
grounds.

Contrary to the ADIOS case, for the ADAF scenario to be
self-consistent, the total kinetic power of the jet/outflow, $L_{\rm
kin}$, should be a negligible fraction of $L_{\rm tot}$.  In order to
verify this, $L_{\rm kin}$ can be estimated making use of the
normalization for the jet kinetic power vs. radio luminosity derived
by~\cite{heinzgrimm}. Obviously, if this number is high enough, the
jet contribution to the flow energetics (and dynamics) can be
negligible with respect to advective cooling.  In~\cite{heinzgrimm}
the radio core emission of three well studied radio galaxies (M87, Per
A and Cyg A) was directly compared to the radio lobe emission, used a
jet calorimeter. They proposed that the jet kinetic power can be
estimated from the core radio luminosity in the following way: $
L_{\rm kin} = 6.2 \times 10^{37} \left(\frac{L_{\rm R}}{10^{30}{\rm
erg}\;{\rm s}^{-1}}\right)^{\frac{1}{1.4-\alpha_r/3}}{\cal W}_{37.8}$
{erg} {sec}$^{-1}$ where $\alpha_r$ is the radio spectral index, and
the parameter ${\cal W}_{37.8}$ carries the (quite large) uncertainty
on the radio galaxy calibration.  For
\sr, assuming a flat radio spectral index $\alpha_r=0$, and for
$L_{\rm R}=7.5 \pm 3.7 \times 10^{26}$ erg sec$^{-1}$, there follows:
$L_{\rm kin} \simeq 3.6 \times 10^{35}{\cal W}_{37.8}$~erg sec$^{-1}$,
or $
\frac{L_{\rm kin}}{L_{\rm tot}} \simeq  0.6 \times {\cal W}_{37.8}~{y}^{-1}~(\eta/0.1)^{-1}$. 
This would mean that the jet/outflow carries a significant amount of
the accretion energy budget away from the system. I so, then any realistic 
accretion flow model for quiescence shall necessarily incorporate the
effects of an outflow both in terms of energetics and dynamics,
effectively ruling out a pure ADAF solution.

Using the estimate for the kinetic power of the jet in quiescence, we
can calculate the total energy carried out by it in between outbursts,
assuming again that the 58 years recurrence time is not overestimated.
We obtain $E_{\rm jet}\simeq 6.6 \times 10^{44} {\cal W}_{37.8}$ erg,
of the same order of the energy released during an
outburst. Interestingly, \cite{mhm99} calculated
the outburst evolution for
\sr with a model accounting for evaporation of the cold outer disc
(but neglecting outflows), and concluded that only about one third of
the mass accreted during quiescence needs to be stored in the disc for
the subsequent outbursting episode.
Under the ADIOS working-hypothesis, the mass flowing from the outer disc does
not reach 
the inner region, but is lost in a outflow. The accretion rate is now a
function of radius:
\begin{equation}
\dot M(R)=\left\{ \begin{array}{ll}
        \dot M_{\rm out},   & R_{\rm out}>R>R_{\rm tr}   \\
        \dot M_{\rm out}(R/R_{\rm tr})^{\alpha}, & R_{\rm tr}>R>R_{\rm
        in}\\
        \end{array}\right.\;
\end{equation}
where we have assumed that mass loss sets in within the truncation
radius $R_{\rm tr}$. Proceeding as before, we can then estimate the
truncation radius for $\dot M_{\rm in}=\dot M(R_{\rm in}$=$3R_{\rm S})$,
being $R_{\rm S}$ the Schwarzschild radius for a 10 \msun BH.
For $\alpha=1$ we obtain $
R_{\rm tr}\simeq 1.8\times 10^4~(y / w)~(\eta/ 0.1)~R_{\rm S}$
or $\simeq 5.4\times 10^{10}~(y/w)~(\eta/0.1)$ cm, 
to be compared with the orbital separation of about 
$3\times 10^{11}$ cm~\cite{gelino01}. 

Assuming that the
the jet/outflow is powered by the mass lost from the accretion flow,
then its total kinetic power $L_{\rm kin}$ is given by
$
L_{\rm kin}=L_{\rm tot} \left(1-\frac{R_{\rm
      in}}{R_{\rm tr}}\right)\simeq L_{\rm tot},
$
i.e. in the ADIOS framework, a dominant fraction of the total accretion power is
channelled into the jet/outflow.} 

In conclusion, by making use of the estimate of the outer accretion
rate of A0620--00 in quiescence and of the jet radiative efficiency
by~\cite{heinzgrimm}, it has been argued that the
outflow kinetic power accounts for a sizable fraction of the accretion
energy budget, and thus must be important with respect to the overall
accretion dynamics of the system~\cite{gallo06}.  On the other hand,
as noted in the case of elliptical galaxies studied with
Chandra~\cite{allen06}, the large jets' mechanical power (compared to
the available fuel supply) suggests that, before being expelled and
carrying away a substantial fraction of the dissipated accretion
power, the accreted mass must flow deep into the hole's potential
well, were high efficiencies can be reached.


\begin{theacknowledgments}
I wish to thank the organizers for putting together, once again, such an 
interesting conference in such a spectacular location.  This work is supported
by NASA through Chandra Postdoctoral Fellowship Award PF5-60037, issued by the
Chandra X-Ray Observatory Center, which is operated by the Smithsonian
Astrophysical Observatory for and on behalf of NASA under contract NAS8-39073.

\end{theacknowledgments}


\begin{thebibliography}{9}

\bibitem{allen06}
Allen S. W. \etal, 2006, MNRAS, 372, 21


\bibitem{bb99}
Blandford R. D., Begelman M. C., 1999, MNRAS, 303, L1

\bibitem{brock}
Brocksopp C. \etal, 2005, MNRAS, 356, 125

\bibitem{cadolle}
Cadolle-Bel M. \etal, 2006, Proceedings of the VI Microquasar
Workshop: Microquasars and beyond, Proceedings of Science,
ed. T. Belloni

\bibitem{chaty}
Chaty S., 2006, Proceedings of the VI Microquasar
Workshop: Microquasars and beyond, Proceedings of Science,
ed. T. Belloni



\bibitem{corbel04}
Corbel S., Fender R. P., Tomsick J. A., Tzioumis A., Tingay S., 2004,
ApJ, 617, 1272

\bibitem{corbel03}
Corbel S., Nowak M., Fender R. P., Tzioumis A. K., Markoff S., 2003, A\&A,
400, 1007





\bibitem{corbel00}
Corbel S. \etal, 2000, A\&A, 359, 251


\bibitem{disalvo01}
Di Salvo T., Done C., Zycki P. T., Burderi L., Robba N. R., 2001, ApJ, 547,
1024

\bibitem{dhawan}
Dhawan V., Mirabel I. F., Rodr\'\i guez L. F., 2000,  ApJ, 543, 373



 	

\bibitem{fender06}
Fender R. P., 2006, in Lewin W. H. G., van der Klis M., eds, Compact
Stellar X-Ray Sources. Cambridge Univ. Press, Cambridge 

\bibitem{fbg}
Fender R. P., Belloni T., Gallo E., 2004, MNRAS, 355, 1105


\bibitem{fb04}
Fender R. P. \& Belloni T. M., 2004, ARA\&A, 42, 317


\bibitem{fgj}
Fender R. P., Gallo E., Jonker P. G., 2003, MNRAS, 343, L99
	






\bibitem{gallo06}
Gallo E. \etal, 2006, MNRAS, 370, 1351


\bibitem{gallo04}
Gallo E., Corbel S, Fender R. P., Maccarone T. J., Tzioumis A. K., 2004,
MNRAS, 347, L52

\bibitem{gfp}
Gallo E., Fender R. P., Pooley G. G., 2003, MNRAS, 344, 60

\bibitem{gelino01}
Gelino D. M., Harrison T. E., Orosz J. E. 2001, AJ, 122, 2668

\bibitem{garcia03}
Garcia M. R., Miller J. M., McClintock J. E., King A. R., Orosz J., 2003, ApJ,
591, 388



\bibitem{heinzgrimm}
Heinz S. \& Grimm H.-J. 2005, ApJ, 633, 384











\bibitem{kong}
Kong A. K. H., McClintock J. E., Garcia M. R., Murray S. S., Barret D., 2002,
ApJ, 570, 277
	
\bibitem{mnw05}
Markoff S., Nowak M. A., Wilms J., 2005, ApJ, 635, 1203

\bibitem{mff}
Markoff S., Falcke H., Fender R., 2001, A\&A, 372, L25

\bibitem{marti02}
Mart\'{i} J., Mirabel I. F., Rodr\'\i guez L. F., Smith I. A., 2002, A\&A,
386, 571 





\bibitem{mhm99}
Meyer-Hofmeister E. \& Meyer F. 1999, A\&A, 348, 154

\bibitem{mccr06}
McClintock J. E., Remillard R. A., 2006, in Lewin W. H. G., van der Klis M.,
eds, Compact Stellar X-Ray Sources. Cambridge Univ. Press, Cambridge

\bibitem{mcc03}
McClintock J. E. \etal, 2003, ApJ, 593, 435

\bibitem{mcc95}
McClintock J. E., Horne K., Remillard R. A. 1995, ApJ, 442, 358


\bibitem{mirabel01}
Mirabel I. F., Dhawan V., Mignani R. P., Rodrigues I., Guglielmetti F., 2001,
Nature, 413, 139

\bibitem{mirabel92}
Mirabel I. F., Rodr\'\i guez L. F., Cordier B., Paul J., Lebrun F., 1992,
Nature, 358, 215
	





	


\bibitem{quataert}
Quataert E., Gruzinov A., 2000, ApJ, 539, 809

\bibitem{ny95}
Narayan R., Yi I. 1995, ApJ, 452, 710 
\bibitem{ny94}
Narayan R., Yi I. 1994, ApJ, 428, L13






\bibitem{rodriguezj}
Rodriguez J. \etal 2006, ApJ submitted (astro-ph/0611341)




\bibitem{ss73}
Shakura N. I., Sunyaev R. A., 1973, A\&A, 24, 337



\bibitem{stirling}
Stirling A. M. \etal, 2001, MNRAS, 327, 1273   





\bibitem{warner}
Warner B. 1987, MNRAS, 227, 23

\bibitem{yuan}
Yuan F., Cui W., Narayan R., 2005, ApJ, 620. 905

\end{thebibliography}
\end{document}